\documentstyle[twocolumn,aps,prl]{revtex}

\begin{document}

\draft
\title{``Critical'' phonons of the supercritical Frenkel-Kontorova
model:\\
renormalization bifurcation diagrams } 
\author{ Jukka A. Ketoja}
\address{Department of Mathematics, P. O. Box 4,
FIN-00014 University of Helsinki, Finland}
\author{Indubala I. Satija\cite{email}}
\address{
 Department of Physics and\\
 Institute of Computational Sciences and Informatics,\\
 George Mason University,\\
 Fairfax, VA 22030, USA}
\date{\today}
\maketitle
\begin{abstract}

The phonon modes of the Frenkel-Kontorova model are studied both at the 
pinning transition as well as in the pinned (cantorus) phase.
We focus on the minimal frequency of the phonon spectrum and
the corresponding generalized eigenfunction. Using an exact
decimation scheme, the eigenfunctions are shown to have 
nontrivial scaling properties not only at the pinning transition point
but also in the cantorus regime. Therefore the phonons defy localization
and remain $critical$ even where the associated
area-preserving map has a positive Lyapunov exponent. In this region,
the critical scaling properties vary continuously and are
described by a line of renormalization limit cycles.
Interesting renormalization bifurcation diagrams
are obtained by monitoring the cycles as the
parameters of the system are varied from an integrable case to the
anti-integrable limit. Both of these limits are described by a
trivial decimation fixed point. Very surprisingly we find 
additional special parameter values in the cantorus regime 
where the renormalization limit
cycle degenerates into the above trivial fixed point. 
At these ``degeneracy points'' the {\it phonon} hull is represented
by an infinite series of step functions. This novel behavior persists
in the extended version of the model containing two harmonics.
Additional richnesses of this extended model are
the one to two-hole transition line, 
characterized by a divergence in the renormalization cycles, 
nonexponentially localized  phonons, and the preservation of critical behavior
all the way upto the anti-integrable limit.


\end{abstract}

\narrowtext

\section{Introduction}

The Frenkel-Kontorova (FK) model is one of the simplest
prototypes for theoretical studies on incommensurate structures in
solid-state physics \cite{A,AL,CF}. The model consists of an infinite
one-dimensional chain of balls in an external sinusoidal potential
with an elastic interaction between nearest neighbors. If the mean
distance between consecutive balls is not in rational proportion to
the periodicity of the sinusoidal potential, the
corresponding state is called incommensurate. Incommensurate
structures appear in many physical systems such as quasicrystals,
two-dimensional electron systems, magnetic superlattices, and
charge-density waves.

Because of the well-known connection to the area-preserving standard
map, the FK model is important also from the nonlinear-dynamics point
of view. An incommensurate ground state of the FK model
corresponds either to an invariant circle or a cantorus (invariant
Cantor set) for the standard map. In the Kolmogorov-Arnol'd-Moser
(KAM) phase, a configuration associated with a
smooth circle can be described by an analytic hull function which maps
a rigid rotation orbit to the configuration. On the other hand,
a cantorus corresponds to a
purely discrete hull function representable as a convergent series of
step functions \cite{Aetal}. Physically, a state with an analytic hull
function is free to slide whereas a discrete hull function implies a
non-zero minimum force (or field) required for depinning.
The pinning transition corresponding to the breakup of an invariant circle
is usually refered as the transition by breaking of analyticity
(TBA).

The linear excitation or the phonon modes of the FK model
describing the physical stability of the
states are affected in an important way by the TBA.
In the KAM phase and at the TBA, it is possible to find a
translational mode with zero frequency (phason mode) supporting
collective transport. The corresponding generalized phonon
eigenfunctions are ``extended'' (quasiperiodic)
in the KAM phase and ``critical'' (neither
extended nor localized) with non-trivial scaling properties
at the TBA. However, not much has been known 
about the eigenfunctions in the cantorus regime
where the frequencies of phonons are bounded away from zero.
In a related quantum problem, describing an electron in a
quasiperiodic potential, a similar transition results in localized
eigenfunctions \cite{KSloc}. On the other hand,
the absence of localization has been proven
for a quasiperiodic potential
taking a finite number of discrete values \cite{DP}.
Although the potential seen by the phonons is purely discrete in the
cantorus regime, the potential still takes an infinite number of
distinct values which leaves the question on the
nature of the eigenfunctions open. 
There are recent speculations by Burkov et al. \cite{BKB}
on the possibility of having
critical phonon modes at band edges of the phonon spectrum.
However, in their numerical study, where the localization
character is infered by computing the participation ratio, it is rather
difficult to make definite predictions.

A great deal of effort has been put on studying
the non-trivial scale invariance observed at the
TBA \cite{CF,dA,GJSS,PA}. 
MacKay \cite{M2} has adjusted his previous renormalization theory for
the critical invariant circles to explain many scaling exponents 
in the neighborhood of the transition.
However, it has not been clear how
to formulate the scaling theory in the supercritical cantorus region
where many relevant
quantities like the actions or phase space scaling constants 
diverge in renormalizing \cite{M1}. Sneddon et al. \cite{SLK} took a step
in this direction by proposing a renormalization theory of the
pinning threshold in the case of infinite viscosity. 
Working in the Fourier basis, they found three-term harmonic expansions for
both the TBA renormalization fixed point and the new
dynamic-threshold fixed point characterizing the depinning transition
at a critical field in the supercritical regime. The latter fixed point
corresponded to vanishing elastic constants.

In this paper, an exact decimation approach \cite{K,KS,KSloc} is used
to study phonons in the FK model both at the TBA and beyond the transition.  
We show that the minimal frequency
phonon mode remains critical throughout the
cantorus regime. Furthermore, the scaling properties of
the corresponding generalized eigenfunctions vary continuously and 
are characterized by a line of limit cycles
of the renormalization equations.
We numerically obtain a full bifurcation diagram of these cycles
as the system parameter is varied upto the anti-integrable
limit where the elastic interaction disappears.
The limit cycle degenerates to a
trivial fixed point at the anti-integrable limit  and at an infinite
sequence of other parameter values approaching the TBA from above.
At this sequence of points, the phonon eigenfunctions are represented
by infinite series of step functions and hence provide novel
behavior in the study of systems
with competing periodicities. The degeneracy points persist in the FK model
with two harmonics and hence are believed to be generic in
one-parameter families of similar models.

We also investigate the scaling properties of
one to two-hole transition\cite{BM2}
for cantori
in a two-harmonic model. Using a multi-precision software
package\cite{B}, we show that 
the transition is signalled by a divergence in one
member of the limit cycle of the renormalization trajectory. However,
the finite members of the limit cycle vary continuously along the
transition line in the two-parameter space.
This transition line is distinguished from the rest of the cantorus
phase by the fact that the phonon eigenfunctions vanish asymptotically.
However, the localization is not exponential as suggested by 
vanishing of the Lyapunov exponent. Another new feature existing only for
a two-hole cantorus is the non-trivial anti-integrable limit for phonons,
i.e. the phonons remain critical all the way upto the anti-integrable limit.

The FK model, its relation to the standard map and the phonon modes describing
the physical stability of the FK model
are briefly reviewed in Section II. Section III explains the
decimation method which has been previously applied to investigate the
critical invariant circles in dissipative systems \cite{K} and
the self-similar wave functions in tight binding models
\cite{KS,KSloc}. In Section IV we give an account of the
numerical methods used in our study. The results of applying the
decimation method to the FK model are reported in Section V.
Analogous results on the extended FK model with two harmonics
are found in Section VI. The main emphasis there is on the one to two-hole
transition. The results and some open problems are discussed in Section VII.
An appendix contains a detailed discussion on what happens to the
phonons approaching the anti-integrable limit.

\section{Generalities}

This section is about the FK model, terminology and basic results of
Aubry's \cite{A,AL} theory.

A configuration of the FK model is defined as the sequence $x_i$,
$i \in Z$, where $x_i$ gives the position of the $i$th
ball. The energy of the configuration is given by
\begin{equation}
W=\sum_i v(x_i ,x_{i+1} ) ,
\end{equation}
where $v$ is the generating function
\begin{eqnarray}
v(x_i ,x_{i+1} ) &=& \frac{1}{2} (x_{i+1} -x_i -a)^2 + V(x_i ) ,\\
V(x) &=& {k\over 4\pi^2} \cos(2\pi x) .
\end{eqnarray}
Because the external potential part $V$ of the energy has the period unity, two
configurations $x$ and $\tilde x$ can be identified if there exist
$m,n \in Z$ such that the transformation
$(T_{mn} x)_i = x_{i+m} -n$ maps $x$ on $\tilde x$. An equilibrium state with
$\partial W / \partial x_i  = 0$ for all $i\in Z$
is obtained by iterating the area-preserving standard map
\begin{equation}
(x_{i+1},y_{i+1})=F(x_i ,y_i )=(x_i +y_{i+1},y_i +V'(x_i)) ,
\end{equation}
where $y_i =x_i -x_{i-1}$.

A {\it minimum energy (m.e.) state} is defined as a configuration $x$ such that
the energy for all other configurations which agree with $x$ outside
a finite range of $i$ is greater than or equal to the energy of $x$.
Every m.e. state is rotationally ordered \cite{AL} which
implies the existence of the mean spacing
of the balls given by the rotation number
\begin{equation}
\sigma = \lim_{N\to \infty} {x_N -x_0 \over N} .
\end{equation}
Incommensurate states correspond to irrational values of $\sigma$. 

One is usually interested in {\it recurrent} m.e. states which are
called {\it ground states} by Aubry \cite{A}.
There exists a closed non-empty
set of recurrent m.e. states of every mean spacing $\sigma \in R$.
If $\sigma$ is irrational, this set corresponds either to an invariant circle
or a cantorus around the cylinder obtained by identifying $x$ and
$x+1$. In other words, the FK model can have an infinite number of ground
states all corresponding to the same invariant circle or cantorus
of the standard map.
An invariant circle can be described by a unique
strictly increasing continuous {\it hull function} $g: R \to R$ so that
\begin{equation}
x_i =g(i\sigma + \phi ), \;\;g(\theta +1)=g(\theta)+1, \label{hull}
\end{equation}
where by varying the phase $\phi$ one can choose the ground state one wants.
There is a similar representation for a cantorus
in terms of either the left- or
right-continuous extension of a unique non-decreasing discontinuous
hull function. 

The {\it physical} ground states are obtained by minimizing the energy
per ball over $\sigma$. For the FK model, the parameter $a$ can be
tuned for each $k$ so that the physical ground states have a desired
mean spacing.

The equation describing the physical stability of the ball configurations
in the FK model is obtained by
linearizing the dynamics 
$d^2 x_i / dt^2  = - \partial W / \partial x_i  $
about an equilibrium configuration. The resulting equation for
the small displacements $\xi_i$ is
\begin{equation}
{d^2 \xi_i \over dt^2 } = - {\partial^2 W \over \partial x_i \partial x_j}
\xi_j 
\end{equation}
which can be written in the compact way
 $d^2 \xi /dt^2 = -D^2 W \xi$. The {\it phonon modes} are solutions
of the form $\xi (t)= exp(i\omega t) \psi$, where 
the (generalized) eigenvector $\psi$ of
the operator $D^2 W$ satisfies the equation  
\begin{equation}
\psi_{i+1} + \psi_{i-1} - [2 + V''(x_i )]
\psi_i = - \omega^2 \psi_i . \label{TBM}
\end{equation}

The spectrum of $D^2 W$, given in terms of the squares of phonon
frequencies $\omega^2$, relates in an important way to the properties
of the ground states of the FK model. First, it
can be proven that the spectrum of $D^2 W$ does not depend on
the phase of an incommensurate ground state in the FK model \cite{Aetal}.
For an invariant circle, the infimum $E_{min}$ of the spectrum
is zero whereas for a uniformly hyperbolic cantorus $E_{min} >0$
\cite{AMB}.
In general, the configurations with $E_{min} > 0$ are physically
at least metastable. Morover,
uniform hyperbolicity implies structural stability, i.e. there cannot
be any bifurcations of the cantorus at the corresponding parameter
values. However, the cantorus can lose uniform hyperbolicity at a 
bifurcation where $E_{min} =0$. Such bifurcations appear in 
extended versions of the FK model containing two or more harmonics
\cite{BM1,BM2,KM,SUW}. 

\section{Phonon decimation}

The phonon equation (\ref{TBM}) can be interpreted as a discrete 
Schr\"odinger equation for the ``wave function'' $\psi_i$ in
the quasiperiodic potential $2+V''(g(i\sigma + \phi))$.
Similar quasiperiodic linear-difference equations 
appear in many other nonlinear dynamics and
solid state problems.  
We have developed a decimation method to deal with such equations
 \cite{K,KS,KSloc}. 
The main advantage of the method is that the scaling properties of
solutions can be described by bounded decimation functions.
The decimation strategy is determined by the underlying ``frequency''
$\sigma$. Our decimation can be implemented for any irrational
$\sigma$ by choosing an arbitrary path in the Farey tree \cite{KS}. 
However, in this paper we will restrict ourselves to one of the noble
rotation numbers, namely, $\sigma =\gamma^{-2}$, where
$\gamma = \frac{1}{2}(1+\sqrt 5)$ is the golden ratio.

In the decimation scheme, all other lattice sites except
those labelled by the Fibonacci numbers $F_n$, $F_{n+1}=F_n +
F_{n-1}$ $(F_0 =0, F_1 =1)$, are decimated out beginning from an
arbitrary initial site $i$.
At the $n^{th}$ step, the decimated phonon equation has the form 
\begin{equation}
f_n(i) \psi(i+F_{n+1})=\psi(i+F_n) + e_n(i) \psi(i). \label{Dec}
\end{equation}
The phonon frequency is simply a parameter in this decimation equation
which therefore should not be considered as an eigenvalue problem.
Using the additive property of the Fibonacci numbers, straightforward
manupulations give the following exact recursion relations for
the decimation functions $e_n$ and $f_n$:
\begin{eqnarray}
e_{n+1} (i) &=& - {A e_n (i) \over 1+Af_n (i)} \label{rec1}\\
f_{n+1} (i) &=& {f_{n-1} (i+F_n) f_n(i+F_n)\over 1+Af_n(i)} \label{rec2}\\
A &=& e_{n-1} (i+F_n) + f_{n-1} (i+F_n)e_n(i+F_n). \nonumber 
\end{eqnarray}
The above recursion relations
can be iterated numerically for an arbitrary number of decimation
steps provided the phonon frequency $\omega$ and the sequence $x_i$,
which set the initial functions $e_2 (i)$ and $f_2 (i)$,
can be calculated with arbitrary precision.

There are a number of reasons why the iteration of the recursion
relations (\ref{rec1}-\ref{rec2}) is superior to the direct iteration of
the phonon equation (\ref{TBM}):\hfill\break
1) Often the iteration of the recursion relations is numerically more stable
than that of Eq. (\ref{TBM}).\hfill\break
2) The possible self-similarity observed by
monitoring the behaviour of $\psi_i$ over the range $i\in [-F_n ,F_n]$
with increasing $n$ can be captured by a simple asymptotic limit cycle
for the decimation functions $e_n$ and $f_n$.\hfill\break
3) An unknown or inaccurate parameter in the equation (\ref{Dec}) and the
asymptotic limit cycle can be simultaneouly determined self-consistently
(see the next section for further details). In other words,
the decimation equations themselves provide a new method
to compute e.g. eigenvalues or phase boundaries upto machine
precision.

The decimation scheme can also be use in
describing the self-similar fluctuations of exponentially localized modes.
Here the localized eigenfunction is written as
$\psi_i = e^{ -\gamma |i|} \eta_i $ which results in the following
equation for $\eta_i$:
\begin{equation}
e^{ -\gamma } \eta_{i+1} + e^{ \gamma}  \eta_{i-1}
 - [2+ V''(x_i )]
\eta_i = - \omega^2 \eta_i
\end{equation}
The above equation can be studied using the decimation scheme
with the inverse localization length
$\gamma$ being an additional parameter. Again $\gamma$
can be determined self-consistently along with the limit cycle.
\cite{footnote1}

Although the above scheme with a discrete lattice index $i$ is sufficient
for the present purpose, we would like to point out that 
an analogous form of the recursion relations in terms of a continuous
variable is needed if one wants to solve 
renormalization limit cycles by the Newton method
\cite{KS,KSloc}.

\section{Numerical Methods}

In this paper, the properties of
the solutions to Eq. (\ref{TBM}) are studied
at the minimal frequency $\omega =\sqrt E_{min}$.
In the KAM phase, the
ground state configurations of the model exhibit a zero frequency phonon.
In this case,
the corresponding eigenfunction can be determined very accurately.
At the TBA, the precision in obtaining the
asymptotic scaling properties is limited by the precision of the
critical $k$ value and the sequence $x_i$. On the other hand,
the iteration of the decimation equations
for an arbitrary value of the nonlinearity parameter $k$ in the cantorus
regime is mainly limited by
the precision of the phonon frequency. However, as pointed out in the
previous section, the existence of asymptotic limit cycles for the decimation
functions can be used to obtain the frequency 
to high precision so that the remaining source of numerical
error is the inaccuracy of the ball positions. This error can be made
very small by approximating a ground state with very long orbit.

There are essentially two steps involved in obtaining the scaling properties
of the phonon eigenmodes. The first step is to obtain 
an estimate for the minimal frequency of the phonon spectrum using an
orbit of moderate size (e.g. 610) in the diagonalization procedure.
The frequency estimate has to be precise enough in order to allow 
so many decimation steps that the asymptotic cycle is already
approached. Usually the diagonalization procedure is the
most CPU time consuming
part in the numerical calculations.
The second part is to approximate the cantorus with a very long orbit
(e.g. of the length 46368) and to improve the frequency estimate to very
high precision so that the recursion relations
(\ref{rec1}-\ref{rec2}) can be iterated  
many times to obtain the asymptotic cycle with several significant digits.

We follow Baesens and MacKay 
(see Appendix A of ref. \cite{BM2}) in calculating an 
initial estimate for $E_{min}$ and the sequence $x_i$
for the decimation equation (\ref{TBM}). 
An invariant circle or a cantorus of rotation number $\gamma^{-2}$
is approximated by periodic orbits
of rotation numbers $F_{n-2}/F_n$. It is sufficient to consider
symmetric periodic orbits which fall into four classes \cite{BM2}:
1) even period with $x_0 =0$, 2) even period with $x_0 =-x_1$, 3)
odd period with $x_0 =0$, and 4) odd period with $x_0 = 1/2$.
These classes are useful in obtaining initial conditions for
the iteration of Eq. (\ref{TBM}) as the phonon modes in each class
are either symmetric or antisymmetric \cite{BM2}.
In the KAM phase and at the TBA, the invariant circle describing the
ball configurations can be approximated by any of the above four classes.
However, a cantorus can be approximated only by class-2 (or class-4 if
the cantorus has only one hole) orbits.
The classes 1 and 3 (and 4 in the region of two holes, see section VI)
 give some homoclinics in
addition to the cantorus so $E_{min}$ can be different for them.
Baesens and MacKay explain how to obtain the orbit Lyapunov
exponent or the residue needed in locating the TBA
and the one to two-hole transition on the parameter axis. 
It should be noted that in the cantorus phase, the standard map
has a positive Lyapunov exponent which near anti-integrable limit
exhibits a logarithmic divergence with the nonlinearity parameter $k$.

One of our main tasks was to decide on whether the eigenfunctions were
exponentially localized or not.
Thouless has derived a formula for the inverse localization length
\cite{S}. However, it practice the estimates obtained with this
formula remain rather crude as it
requires considering differences between all eigenvalues.
Initially, we carried out the decimation varying the value of
the inverse localization length $\gamma$ but we found that only
$\gamma =0$ showed convergence towards
a non-trivial asymptotic $p$-cycle for the decimation functions at
$\omega = \sqrt E_{min}$. Assuming $\gamma=0$,
the value of $E_{min}$ was then sequentially improved with the secant method by
 requiring
that the decimation functions asymptotically converge to a period $p$
limit cycle. In particular, we used
 $f_{n} (0) =f_{n-p}(0)$ with e.g. $n=17,20,23$,
where the cantorus was approximated
with an orbit of the length $F_{23}$=$46368$.
Table I compares $E_{min}$ obtained in this way at $k=2$ with
diagonalization results.  One should note that the above self-consistent 
calculation of $E_{min}$ is not only more accurate but also takes only
a small fraction  of the CPU time 
needed for the diagonalization of a matrix of the size 10946/2.
All the computations  were done
using quadruple precision.

The critical line for
the one to two-hole transition was determined by the
symmetry breaking bifurcations as in ref. \cite{BM2}.
Table II  lists the bifurcation points of 
periodic orbits along the line $k_2=4k_1$.
It can be seen in this table that due to superexponential convergence,
all decimals of the quadruple precision are already exhausted  at the
orbit length 377. 
This size is highly inadequate because it limits the
maximum number of decimation iterations to be 14. With $14$ iterations, we are
still in the transient regime and the existence of a limit cycle for the
decimation functions is far from obvious. Moreover, the nature of
phason eigenfunction or the fate of the orbit Lyapunov exponent cannot
be decided on such a short orbit. 
To overcome this problem, we used
a multi-precision software package \cite{B} to compute the orbit and
the phonon eigenfunction along the one to two-hole transition line
(see Section VI for further details).

\section{Renormalization bifurcation diagram}

The two-step numerical procedure discussed in the previous section
was applied to study the scaling properties of the phonon eigenfunctions
for a multitude of values of $k$ starting from the KAM phase upto the
{\it anti-integrable limit} $k\to \infty$. This limit
corresponds to vanishing elastic interaction \cite{AA} which implies that
for a m.e. state all balls lie at the potential energy minimum $x=1/2$.
By introducing the parameter $r=2\arctan(k) /\pi$ one can map
the whole $k$-axis to the $r$-interval $[0,1]$, where $r=0$ and $r=1$
are the integrable and anti-integrable limits, respectively.

In the KAM region,
where the invariant circle is smooth, one obtains the same
trivial fixed point for all classes of approximating orbits.
It can be easily solved
from the fixed point equation: $f_n (i) \equiv \sigma$ and $e_n (i) \equiv
-\sigma^2$. It describes an extended quasiperiodic phonon eigenfunction.

In general, choosing the class fixes the phase $\phi$
in Eq. (\ref{hull}) and thus different classes
can lead to different decimation cycles.
This is the case at the TBA, where the class 1 and 3 orbits having $x_0$
on the dominant symmetry line lead to a nontrivial universal fixed point
\cite{K} whereas the class 2 and 4 orbits correspond to
3-cycles.
In Fig. 1 we plot the critical phason ($\omega =0$) mode $\psi_i$  obtained by
approximating the invariant circle by the class 1 or 3 orbits which
causes the main peak to be located at $i=0$.
The non-trivial fixed point and the corresponding self-similarity
clearly distinguishes the TBA from the KAM regime.

In the cantorus regime, both classes 2 and 4 lead to the
same decimation 6-cycle for fixed $k$. 
Fig. 2(a) shows the variation of the 6-cycle 
as we change the parameter $r$ from the KAM region
to the anti-integrable limit.
An interesting feature of the diagram is the infinite number of
{\it loops} separated by {\it degeneracy points}
where the decimation functions approach asymptotically the same trivial limit
cycle as in the KAM phase. Coming down from the anti-integrable limit,
the first such degeneracy is found at $r=0.67052674...$. Although the
decimation functions come close to the trivial
fixed point already near $r=0.84$, they behave
parabolically in this region forming an avoided crossing. As shown in the
blowup of Fig. 2(b),
the behaviour is very different around $r=0.67052674$ where
the curves behave linearly as a function of the parameter $r$ all
passing through the inverse golden mean at the same point.
Table III gives a list of degeneracy points. We have checked that they
all correspond to a real crossing of the curves in the bifurcation diagram.
We conjecture that there are an infinite number of such points
accumulating at the critical parameter value for the TBA. 

As can be seen in Table III, the separation between subsequent
degeneracy points appear to asymptotically
scale by the universal scaling constant
$\delta=1.62795...$ for the TBA \cite{M1,M2}.
This suggests that the observed loop
pattern is related to the renormalization dynamics along
the unstable manifold of the TBA fixed point.
This in turn causes one to think that there should be some
property of the configurations which makes the degeneracy points
special. If such a property exists, it has to be rather subtle as
the orbit hull function is given by qualitatively the same kind of
series of step functions through the whole cantorus region.

An important property associated with the degeneracy points is
revealed by plotting the {\it phonon hull}, i.e. $\psi_i$ versus $\{
i\sigma \}$
where $\{ i\sigma \}$ is the fractional part of $i\sigma$.
Fig. 3(a) shows this plot
at one of the degeneracy points. The resulting graph appears to be
represented by an infinite series of step functions. Similar plots are
obtained at other degeneracy points. 
In other words, the phonon
hull shows similar characteristics as the orbit hull function
$g$ at these points.
 It is interesting to note that this is not only true at the degeneracy points
but also in the KAM phase where both the orbit and phonon hull functions
are smooth. This suggests that trivial asymptotic decimation
functions in general
signify qualitatively equivalent behavior in the quasiperiodic
``potential'' term of the linear difference equation and in the corresponding
(generalized) eigenfunction.

In the cantorus regime outside the degeneracy points, the
phonon hull consists of a fractal set of points resembling the phonon
hull function at the TBA. Fig. 3(b) shows how the character of the
phonon hull changes as we move away from the degeneracy points.
The existence of a decimation limit cycle implies self-similarity of
the phonon eigenfunction. This self-similarity is clearly seen after
adjusting the phase factor so that the eigenfunctions
remain bounded asymptotically \cite{footnote2}.
Interestingly, the resulting decimation trajectories converge on a 4-cycle.
The 4-cycle appears to be more
fundamental (and interesting because it breaks the number theoretic
even-odd 3-cycle) than the 6-cycle. It means there is a
point on the cantorus around which the phonon mode is self-similar
although none of the approximating orbits pass through it.
However, since the qualitative features of the bifurcation diagram
(loops and degeneracy points) are similar using the 4 or 6-cycle,
for simplicity, our bifurcation diagrams are shown with 6-cycle.

\section{Two-harmonic model and the one to two-hole transition}

Next we study the phonons and the associated renormalization
bifurcation diagram
in the two-harmonic FK model with potential
\begin{equation}
V(x) = {k_1\over 4\pi^2} \cos(2\pi x) +{k_2\over 16\pi^2} \cos(4\pi x)
. \label{2h}
\end{equation}
Our motivation for this study is two-fold: firstly, to see
if loops and degeneracy points are generic in
such models and secondly, to obtain the renormalization bifurcation diagram
near  the one to two-hole transition for cantori
studied recently by Baesens and MacKay \cite{BM2}. 

As stated in their paper, a {\it gap} in a cantorus
is a pair of distinct points
of the cantorus whose forward and backward orbits converge together.
An orbit of gaps is called a {\it hole}. In the standard map with single
well potential, the cantorus is uniquely determined by the
rotation number (if there is no invariant circle with that rotation
number)  and contains only one hole. However,
in the two-harmonic model
the cantori for a given rotation number are parametrized by
$\alpha \in [0,1]$ near a non-degenerate double-well anti-integrable
limit with $|k_1 | < |k_2 |$ \cite{BM1}. Here $\alpha$ 
gives the fraction of the natural measure of the cantorus in one of
the wells. In other words, there are
uncountably many cantori of each rotation number.  They have
generically two holes although there is a dense set of values of
$\alpha$ corresponding to only one hole. Using symmetry properties it
can be argued for $k_2 > 0$ that the m.e. cantorus has an
equal fraction of points
at both wells near the anti-integrable limit \cite{BM1}.
 Thus Baesens and MacKay \cite{BM2} focus 
on $\alpha =1/2$. Numerically they observe a critical line
in the $k_1-k_2$ parameter space corresponding to the one to
two-hole transition for
the symmetric cantorus with $E_{min} =0$.

Fig. 4(a) shows the renormalization bifurcation diagram obtained by
varying the parameter 
\begin{equation}
r={2\over \pi} \arctan \sqrt{k_1^2 + k_2^2 }
\end{equation}
along the line $k_2=4k_1$. We followed the same numerical
methods as in the previous section and approximated the cantorus
by class-2 orbits (the class-4 orbits give a homoclinic
in addition to the cantorus in the region of two holes).
The figure clearly shows the existence of loops and degeneracy points
near the TBA thus establishing the result that their existence
is generic in one-parameter systems.

The transition from one to two holes is signalled by a divergence
in one of the six members of the limit cycle. The other members of the
limit cycle are found to remain finite. Furthermore, the bifurcation diagram
in the neighborhood of the one to two-hole transition exhibit a rather
strange pattern due to the fact that various members of the limit
cycle cross each other as the parameter $r$ is varied. 
However, the
limit cycle retains the period six at these crossing points.
The crossing pattern becomes more and more complicated near the
1-2 hole transition. There can be a multitude of crossings in a very
narrow parameter interval which makes the behaviour of the curves 
look discontinuous at some points. However, blowups reveal that
the curves are in fact smooth (see Fig. 4(b)).

We have also studied the critical line corresponding to the
hole transition in the $k_1-k_2$ space. Our numerical results strengthen
the previous conjecture \cite{BM2} that the Lyapunov exponent of the
extended standard map vanishes along the hole transition (see Table IV).
The decimation functions obtained with class 2 orbits
converge on a 6-cycle with one 
member of the cycle diverging while the remaing elements of
the 6-cycle remain bounded and vary continuously along the transition
line. Numerically the phason mode is found to be peaked around the
lattice site $i_c$ corresponding to $x_{i_c}$ which is closest to $1/2$.
Therefore, on the transition line it is natural to look at class-4 orbits
because they have the $0$th point exactly at $1/2$.
However, we find no non-trivial limit cycle with the class-4 orbits
reflecting the fact that the phason mode vanishes asymptotically
as $i\to \pm \infty$. This localization is non-exponential with
the inverse localization length $\gamma =0$. In Fig. 5 we plot the
phason mode in both linear-log and log-log scales. These plots
show that the localization is slower
than exponential but faster than geometric.
 To our knowledge, this peculiar behaviour of the
eigenfunctions
has not been observed before.

Another unique aspect of the two-harmonic model is the fact that
it is possible to show the existence of critical phonon modes 
in the anti-integrable limit. This
is clearly seen in Fig. 4(a) where the phonon eigenfunctions are
characterized by a non-trivial limit cycle even approaching $r=1$.
As explained in Appendix, for a generic symmetric two-well potential
the universality class in the anti-integrable limit
is determined by $\sigma$, $\alpha$, and two
additional parameters setting the location of the minima and the ratio
of the third and second derivative of the potential at the minima.
For the potential (\ref{2h}) the latter two parameters are both fixed
by the ratio $k2/k1$. 

\section{Discussion}

Based on exact renormalization methods which can be implemented to
very high precision on the computer, we have shown that the 
phonon eigenfunction at the minimal frequency is self-similar
with scaling properties that vary continuously within the cantorus regime.
A novel feature of our study is the existence of a sequence of points
where the phonon eigenfunctions are represented by series of step functions.
An important aspect of the work described here is the new method 
to compute eigenvalues.
The determination of the bifurcation diagram near the TBA
was a challenging numerical problem and would have been impossible
without the self-consistent approach provided by the decimation
scheme.

It is interesting to compare our results with those for a related electron
problem \cite{KS,KSloc}.
The TBA in the FK model is somewhat analogous to the metal-insulator 
transition for two-dimensional electrons due to the formal similarity
between the phonon equation and the 
corresponding equation for the electron wave functions.
However, the innocent looking difference in the quasiperiodic
potential leads to significantly different behaviour beyond the transition. 
In the ``insulating'' phase of the electron problem,
the wave functions are exponentially decaying with
self-similar fluctuations characterized by a unique renormalization fixed
point. In contrast, the phonon eigenfunctions in the cantorus regime
exhibit self-similarity
without exponentially decaying envelope and can be 
described by a line of renormalization limit cycles.
Furthermore, the cantorus phase contains 
an infinite number of special points where the
renormalization flow converges to the same trivial fixed point
as in the KAM phase or in the ``metallic'' phase of the electron
problem. The anti-integrable limit is trivial for the standard FK
model but non-trivial in the region of a two-hole cantorus.
In comparison, the
counterpart of the anti-integrable limit
in the electron problem, the strong coupling limit
is always highly nontrivial.

The above described degeneracy points remain one of the main
mysteries of our discoveries.
In particular, why do the degeneracy points appear in the cantorus
phase  starting arbitrarily 
close to the TBA and why do
they disappear beyond a certain value of the nonlinearity
parameter?
We have tried to correlate them with some special characteristics of the ball
configurations of the FK model but without success.
Furthermore, the complexity of the renormalization flow near the
one to two-hole
transition is very puzzling and devoids of any explanation.

Our results have  significance in the theory of an almost
periodic eigenvalue problem. Degeneracy points are the first example
in this class of problems whose solutions can be represented by
an infinite series of step functions.
Another curious feature are the non-exponentially decaying
eigenfunctions at the one to two-hole transition in an extended
FK model.

In this paper, we have explored only the parameter region
$k_2 > 0$ of the extended FK model.
However, there is a really rich pattern of cantorus
bifurcations for $k_2 < 0$ \cite{BM1,KM}. In particular,
Schellnhuber et al. \cite{SUW} have reported exponential localization
of the phason mode at a boundary of metastability in an extended version
of the FK model containing three harmonics. The metastability is
related to vanishing of the gap of the phonon spectrum which in that case
signals annihilation of the cantorus configuration.
It would be interesting to apply the decimation method at this
and other similar bifurcations and determine the scaling properties
of the exponentially localized eigenfunctions.

One of the most intriguing features of studying the phonons in the FK model 
is that various different
subjects like condensed matter physics, nonlinear dynamics, 
almost periodic Schr\"odinger operators, and renormalization theory
are drawn together. 
We hope that the results of this paper motivate
further research in each of these fields.

\acknowledgements

We would like to thank J.C. Chaves for his help with multiprecision
software.
The research of IIS is supported by a grant from National Science
Foundation DMR~093296. JAK would like to acknowledge the support from
the Niilo Helander Foundation.
JAK is also grateful for the hospitality
during his visit to the George Mason University. 

\appendix
\section*{Phonons in the anti-integrable limit}

Here we explain the origin of the non-trivial critical
behavior of the phonons for the double-well
anti-integrable limit (AIL). This will account for the
differences at
$r=1$ in Fig. 2(a) and the Fig. 4(a).
In particular, we derive an equation for the critical phonons corresponding
to a two-hole cantorus approaching the AIL.

The key point of the derivation is the fact that
the limiting form of the potential seen by the phonons approaching the
AIL is determined not by the locations of the balls at the
AIL (which gives a constant contribution) but by the small
deviation in the ball positions from those at the AIL.

It is useful to multiply the phonon equation (\ref{TBM})
by the factor
\begin{equation}
t={1\over \sqrt{k_1^2 +k_2^2}}  
\end{equation}
so that all terms in the equation remain bounded in the AIL $t \to 0$:
\begin{equation}
t\psi_{i+1} + t\psi_{i-1} -[2t + U ''(x_i )] \psi_i = -\bar{\omega}^2 
\psi_i , \label{nTBM}
\end{equation}
where $U (x) =tV(x)$ and $\bar{\omega}^2 = t\omega^2$.
It is important that $U$ does not depend on the value of $t$ if the
ratio $k_2 /k_1$ is fixed.
We now expand $U(x)$
around the ball position at the AIL which we denote as $e_i$. Since
$U '(e_i )=0$, we have
\begin{equation}
U'(x_i) = U''(e_i)\xi_i + O(\xi_i^2 ),
\end{equation} 
 where $\xi_i =(x_i -e_i)$ is the small
deviation in the ball configuration from that of the AIL. Using the equation
\begin{equation}
U'(x_i )= t(x_{i+1} -2x_i + x_{i-1} )
\end{equation}
we notice that $\xi_i$ changes linearly with $t$ near the AIL.
Therefore, upto the first order in $t$, we have
\begin{equation}
\xi_i = {t d_i \over U''(e_i )},
\end{equation}
where $d_i = e_{i+1} -2e_i + e_{i-1}$. The next step is to put this
together with the expansion
\begin{equation}
U''(x_i )= U''(e_i ) + U'''(e_i ) \xi_i + O(\xi_i^2)
\end{equation}
into Eq. (\ref{nTBM}).
Because $U '' (e_i )$ is equal for all $i$ if $k_2 > 0$, it is
possible to match the zeroth and first order terms in $t$ so that one
obtains  the asymptotic equation
\begin{eqnarray}
\psi_{i+1} + \psi_{i-1} + V_{AIL} (i) \psi_i = (2+C) \psi_i , \label{AIP}
\end{eqnarray}
where
$C$ is the limit of $-\omega^2 +U''(e_i )/t $ as $t\to 0$ and
\begin{equation}
V_{AIL} (i) = - {U'''(e_i ) d_i \over U''(e_i )} .
\end{equation}

In the case of a single well, the potential is even around the minimum
and the resulting asymptotic phonon equation becomes simply 
\begin{eqnarray}
\psi_{i+1} + \psi_{i-1}  = (2+C) \psi_i .
\end{eqnarray}
$E_{min}$ corresponds to the supremum of the spectrum which in this
case is $2$, i.e. $C=0$.

In the case of two wells for $|k_1|<k_2 $
a straightforward calculation shows that
\begin{equation}
{U''' (e_i ) \over U'' (e_i )} = \pm {6\pi \over \sqrt{(k_2 /k_1 )^2
-1}} \label{V23} ,
\end{equation}
where the $+$ ($-$) sign corresponds to
$\{ e_i \} > 1/2$ ($\{ e_i \} < 1/2$).    
Let us next study the possible values of $d_i$
assuming that $\sigma = (3-\sqrt 5)/2$. The hull function which gives the
set $e_i$ at the AIL reads \cite{BM1,BM2}
\begin{equation}
g_e (\theta )= \cases{ x_m + Int(\theta ), 0<\{\theta \} < \alpha \cr
1-x_m + Int(\theta ), \alpha < \{ \theta \} < 1 } ,
\end{equation}
where $x_m$ is the minimum in $(0,1/2)$ satisfying $\cos(2\pi x_m
)=-k_1 /k_2$. In the symmetric case $\alpha =1/2$ it is easy to
see that $d_i$ can have only the values
\begin{equation}
d_i= \cases{ \pm 2x_m , \theta_i \in \{(0,1/2-\sigma), 
(1/2+\sigma ,1)\} \cr
\pm (4x_m -1), \theta_i \in \{(1/2-\sigma,\sigma), 
(1-\sigma,1/2+\sigma)\}  \} \cr
\pm (2x_m -1), \theta_i \in (\sigma , 1-\sigma) } ,
\end{equation}
where $\theta_i = i\sigma +\phi$ and the $\pm$ sign cancels the one coming
from Eq. (\ref{V23}).
Because the potential $V_{AIL}$ takes only three different values, the 
generalized eigenfunctions cannot be localized \cite{DP}. 
$E_{min}$ corresponds to the supremum of the spectrum of
Eq. (\ref{AIP}). The class 2 orbit corresponds to choosing the phase 
$\phi =-\sigma /2$ in the hull function \cite{BM2}. Decimation
with this phase leads to the non-trivial 6-cycle found by
extrapolating the curves of a bifurcation diagram to the AIL.

It is interesting to note that $V_{AIL}$ is determined solely by
$\sigma$, $\alpha$, $x_m$, and a multiplying factor.
Therefore, other similar two-well potentials for which these
parameter agree would give rise to the same universality class in
the AIL.

\begin{figure}
\caption{Universal phason mode at the TBA with $x_0 =0$.} 
\label{fig1}
\end{figure}

\begin{figure}
\caption{(a) Renormalization bifurcation diagram showing the
asymptotic behaviour of the decimation function $f$
 as a function of the parameter $r=2\arctan(k) /\pi$.
 At each parameter value, $f_n
(0)$ is plotted for at least $(p+2)$ different decimation levels $n$
which shows the accuracy of the limit cycle (of length $p$) obtained.
The 3-cycle at the TBA is plotted with darker points.
The curves behave parabolically
near $r=0.84$ and do not intersect there.
(b) Blowup around the first degeneracy point where the branches of the
6-cycle meet at an intersection point.} 
\label{fig2}
\end{figure}

\begin{figure}
\caption{Phonon hull at the second degeneracy point $r=0.62221...$
(a) and away from it at $r=0.62$ (b).}
\label{fig3}
\end{figure}

\begin{figure}
\caption{Renormalization bifurcation diagram for the
extended FK-model with $k_2 =4k_1$. (a) shows the elements of the
limit  cycle that
remain bounded by $2$ near the one to two-hole transition
and (b) shows an example of how
complicatedly the curves can cross one another in a very narrow
parameter interval.
For an analogous configuration bifurcation diagram see Fig. 1 of ref. 10.} 
\label{fig4}
\end{figure}

\begin{figure}
\caption{Phason eigenfunction with $x_0 =1/2$ at the one to two-hole transition
plotted in linear-log (a) and log-log (b) scales. These plots show that
the localization is between exponential and power-law. } 
\label{fig5}
\end{figure}

\begin{table}
\caption{Estimates for the infimum of the phonon spectrum $E_{min}$ at
$k=2$. The values in the third column were obtained by diagonalizing
a matrix of size $F_n /2$ as described in ref. 10.
The values in the last column were determined by requiring that
$f_n (0)=f_{n-6} (0)$ where the sequence $x_i$ was approximated by
the class-2 orbit of the length 46368.}
\begin{tabular}{cccc}
 & \\
$n$ & $F_n$ & $E_{min}$ (diag.) & $E_{min}$ (dec.) \\
 & \\
\tableline
 & \\
14 & 610 & 0.5935392137880 & 0.5935394324321 \\
 & \\
17 & 2584 & 0.5935394985221 & 0.5935394834426 \\ 
 & \\
20 & 10946 & 0.5935394812326 & 0.5935394820297 \\
 & \\
23 & 46368 & & 0.5935394820545 \\ 
 & \\
\end{tabular}
\label{table1}
\end{table}

\begin{table}
\caption{Symmetry breaking bifurcations of class 4 orbits along the line
$k_2=4k_1$.  The bifurcation points have been calculated using 
multiprecision software based on doubly precision arithmetics.
We have checked consistence with a direct calculation using quadruple
precision. The bifurcation points
converge fast to the critical parameter value for the one to
two-hole transition. 
Note that the corresponding value 
0.429439737992501223898752836653609 obtained in ref. 10
appears to be accurate only upto double precision. }
\begin{tabular}{cc}
 & \\ 
 $F_n$  &  $k_1$ \\
 & \\ 
\tableline 
 & \\
377 &0.4294397379925012239000251474523975392588953 \\
    &5245566429536, \hfill \\
 & \\
987 &0.4294397379925012239000251474523975392588953 \\
    &7931282057125515563523777678 \\
    &714122115567680849012510667905, \\
 & \\
1597 &0.4294397379925012239000251474523975392588953 \\
     &7931282057125515563523777678 \\
     &9131402934114670775341029280082904661719, \\
 & \\ 
\end{tabular}
\label{table2}
\end{table}

\begin{table}
\caption{Degeneracy points approaching the critical parameter value
$r_c$ from above and their scaling.}
\begin{tabular}{ccc}
 & & \\
$n$ &  $r_n$ & $(r_{n-1} -r_c)/(r_n -r_c)$ \\
 & & \\
\tableline
 & & \\
0 & 1.00000000 & \\
 & & \\
1 & 0.67052674 & 2.8336 \\
 & & \\
2 & 0.62221221 & 1.3678 \\
 & & \\
3 & 0.56379391 & 1.8008 \\
 & & \\
4 & 0.53923590 & 1.5075 \\ 
 & & \\
5 & 0.51909955 & 1.7126 \\
 & & \\
6 & 0.50876597 & 1.5765 \\
 & & \\
7 & 0.50162025 & 1.6630 \\
 & & \\
8 & 0.49754645 & 1.6076 \\
 & & \\
\end{tabular}
\label{table3}
\end{table}

\begin{table}
\caption{Lyapunov exponent $\lambda$ of the $F_{n-2}/F_n$ orbit (class
2) at the one to two-hole
transition point with $k_2 =4 k_1$. The transition
point was approximated by the last parameter value in Table II.
The symmetry breaking points of the odd cycles converge faster than
those of the even cycles which justifies the inclusion of the Lyapunov
exponent of the 2584-orbit in this table.}
\begin{tabular}{cc}
 & \\
$F_n$ & $\lambda$ \\
 & \\
\tableline
 & \\
34 &  0.38572449 \\
 & \\
144 & 0.20463797 \\
 & \\
610 & 0.10618378 \\
 & \\
2584 & 0.05450141 \\ 
 & \\
\end{tabular}
\label{table4}
\end{table}


\begin{references}
\bibitem{footnote1} If the decimation was carried for $\psi_i$, the
inverse localization length would be determined by the decay rate
of $e_n$, i.e. 
$\gamma = \lim_{n \to \infty} \ln (e_n (0))/F_n$.
\bibitem{footnote2}
Although the inverse localization length $\gamma$ is zero,
the phonon eigenfunctions corresponding to the nontrivial 6-cycles 
diverge very slowly as $i \to \pm \infty$
at other than the
Fibonacci sites.
There are general results on almost periodic Mathieu operators
\cite{S} which claim the existence of a critical phase $\phi_c$
for which the generalized eigenfunction should be bounded.
Numerically we find the point $x_c =g(\phi_c )$ on
the cantorus by monitoring the largest
peak of the phonon mode in a finite region $-F_n < i  < F_n $.
Assuming it takes place at the lattice site $i(n)$,
$x_c$ can approximated by $x_{i(n)}$.
Increasing $n$ we can get arbitrarily close to $x_c$
and using lattice translations we can carry out the decimation
with $x_0 \approx x_c$. The eigenfunction is
not symmetric around this point. See ref. \cite{KS} for analogous results on
the electron tight binding model.
\end{references}
\end{document}